\documentclass{article}

\begin{document}

\centerline{\Large The Necessary and Sufficient Conditions of Separability} %
\centerline{\Large for Multipartite Pure States }
\footnote{The paper was supported by NSFC(Grant No. 60433050), the fundamental research fund of Tsinghua university NO: JC2003043 and partially by the state key lab. of intelligence technology and system}

\centerline{Dafa Li$^{1,*}$, Xiangrong Li$^{2}$, Hongtao Huang$^{3}$, Xinxin
Li$^{4}$ }
\centerline{$^{1}$ Dept of mathematical sciences, Tsinghua
University, Beijing 100084 CHINA} \centerline{email:dli@math.tsinghua.edu.cn}
\centerline{$^{2}$ Department of Mathematics, University of California,
Irvine, CA 92697-3875, USA}

\centerline{$^{3}$ Electrical Engineering and Computer Science Department} %
\centerline{University of Michigan, Ann Arbor, MI 48109, USA} %
\centerline{$^{4}$ Dept. of computer science, Wayne State University,
Detroit, MI 48202, USA}

Abstract

In this paper we present the necessary and sufficient conditions of
separability for multipartite pure states. These conditions are very simple,
and they don't require Schmidt decomposition or tracing out operations. We
also give a necessary condition for a local unitary equivalence class for a
bipartite system in terms of the determinant of the matrix of amplitudes and
explore a variance as a measure of entanglement for multipartite pure states.



\section{Introduction:}

Notation: $M^{+}$ is the complex conjugate of transpose of $M$.

Let $|\psi \rangle $ and $|\phi \rangle $ be two pure states of a composite
system $AB$ possessed by both Alice and Bob, where system $A$ ($B$) is
called Alice's (Bob's) system. By Nielsen's notation $|\psi \rangle $ $\sim $
$|\phi \rangle $ if and only if $|\psi \rangle $ \ and $|\phi \rangle $ are
locally unitarily equivalent \cite{Nielsen99}. Let $\rho _{\psi }^{A}$ and $%
\rho _{\phi }^{A}$ be the states of Alice's system. It is known that $|\psi
\rangle $ $\sim $ $|\phi \rangle $ \ if and only if $\rho _{\psi }^{A}$ and $%
\rho _{\phi }^{A}$ have the same spectrum of eigenvalues \cite{Nielsen99} %
\cite{Peres}. A pure state is separable if and only if it can be written as
a tensor product of states of different subsystems. It is also known that a
state $|\psi \rangle $ of a bipartite system is separable if and only if it
has Schmidt number 1 \cite{Nielsen00}. Clearly it is essential to do Schmidt
decomposition to find the eigenvalues of $\rho _{\psi }^{A}$ and $\rho
_{\phi }^{A}$. To obtain a Schmidt decomposition of a pure state $|\psi
\rangle $, we need to compute (1) the density operator $\rho _{\psi }^{AB}$;
(2) the reduced density operator $\rho _{\psi }^{A}$ for system $A$; (3) the
eigenvalues of $\rho _{\psi }^{A}$. However it is hard to compute roots of a
characteristic polynomial of high degree.

Peres presented a necessary and sufficient condition for the occurrence of
Schmidt decomposition for a tripartite pure state \cite{Peres95}\ and showed
that the positivity of the partial transpose of a density matrix is a
necessary condition for separability \cite{Peres96}. Thapliyal showed that a
multipartite pure state is Schmidt decomposable if and only if the density
matrices obtained by tracing out any party are separable \cite{Thapliyal99}.
In \cite{Grassl} the local invariants of quantum-bit systems were
investigated. In \cite{Sudbery00}\cite{Sudbery01} the local symmetry
properties and local invariants of pure three-qubit states were discussed,
respectively. In \cite{Acin} the classification of three-qubit states was
given. Bennett reported measures of multipartite pure-state entanglement in %
\cite{Bennett00}. Meyer and Wallach \cite{Meyer}\ proposed a measure of $n-$%
qubit pure-state entanglement. Nielsen used the majorization of the
eigenvalues of the reduced density operators of a composite system $AB$ to
describe the equivalence class under LOCC transformations.

For a multi ($n$)$-$partite system, in this paper we illustrate the reduced
density operators obtained by tracing out the $ith$ subsystem $\rho
^{12...(i-1)(i+1)...n}=tr_{i}(\rho ^{12...n})=M_{i}M_{i}^{+}$, where $%
i=1,2,...,n$\ and $M_{i}$ are the $d^{n-1}\times d$\ matrices, of which
every entry is an amplitude of the state in question. For a bipartite system
$AB$, the reduced density operator $\rho _{\psi }^{A}$ ($\rho _{\psi }^{B}$)$%
=MM^{+}$, where $M$ is the matrix of the amplitudes. Hence $\det (\rho
_{\psi }^{A})=|\det (M)|^{2}$. However, for a multi ($n$)$-$partite system, $%
M_{i}$ are not square. In section 2, we present a necessary and sufficient
condition for separability for a bipartite system in terms of the
determinants of all the $2\times 2$\ submatrices of the matrix of the
amplitudes. Section 3 contains three versions of the necessary and
sufficient separability criterion for a $n-$qubit system. Section 4 is
devoted to study the separability of multipartite pure states, and two
versions of the necessary and sufficient separability criterion are
proposed. Section 5 gives a simple necessary criterion for $|\psi \rangle $ $%
\sim $ $|\phi \rangle $ for a bipartite system. Section 6 suggests an
intuitive measure of multipartite pure-state entanglement.

\section{The separability for a bipartite system}

Let $|\psi \rangle $ be a pure state of a composite system $AB$ possessed by
both Alice and Bob. In this section we give a simple and intuitive criterion
for the separability. Let $|i\rangle $ ($|j\rangle $) be the orthonormal
basis for system $A$ ($B$). Then we can write $|\psi \rangle
=\sum_{i,j}a_{ij}|i\rangle |j\rangle $, where $%
\sum_{ij=0}^{n-1}|a_{ij}|^{2}=1$. Let $M$ $=(a_{ij})_{n\times n}$ be the
matrix of the amplitudes of $|\psi \rangle $. Then the criterion for the
separability is as follows.

$|\psi \rangle $\textbf{\ }is separable if and only if the determinants of
all the $2\times 2$\ submatrices of $M$\ are zero.\textbf{\ }

This criterion for the separability avoids Schmidt decomposition. To compute
the determinants, it needs $n^{2}(n-1)^{2}/2$ multiplication operations and $%
n^{2}(n-1)^{2}/4$ minus operations.

Proof. Suppose that systems $A$ and $B$ have the same dimension $n$.\ By
definition, $|\psi \rangle $ is separable if and only if we can write $|\psi
\rangle =(\sum_{i=0}^{n-1}x_{i}|i\rangle )\otimes
(\sum_{j=0}^{n-1}y_{j}|j\rangle )$, where $\sum_{i=0}^{n-1}|x_{i}|^{2}=1$
and $=\sum_{j=0}^{n-1}|y_{j}|^{2}=1$. By tensor product $|\psi \rangle
=\sum_{i,j=0}^{n-1}x_{i}y_{j}|i\rangle |j\rangle $. It means that $|\psi
\rangle $ is separable if and only if $x_{i}y_{j}=a_{ij}$, $%
i,j=0,1,...,(n-1)......(1)$. Let $m=\left(
\begin{tabular}{cc}
$a_{il}$ & $a_{ik}$ \\
$a_{jl}$ & $a_{jk}$%
\end{tabular}%
\right) $ be any $2\times 2$ submatrix of $M$. It is easy to check $\det
(m)=a_{il}a_{jk}-a_{ik}a_{jl}=x_{i}y_{l}x_{j}y_{k}-x_{i}y_{k}x_{j}y_{l}=0$.
Therefore if $|\psi \rangle $ is separable then the determinants of all the $%
2\times 2$ submatrices of $M$\ are zero.

Conversely, suppose that the determinants of all the $2\times 2$ submatrices
of $M$ are zero. We can write $M$ in the block form, $M=\left(
\begin{tabular}{c}
$A_{0}$ \\
$A_{1}$ \\
$\vdots $ \\
$A_{n-1}$%
\end{tabular}%
\right) =(B_{0},B_{1},...,B_{n-1})$, where $A_{i}$ is the $ith$ row and $%
B_{i}$ is the $ith$ column of $M$, respectively, $i=0,1,...,(n-1)$. Let $%
\left| x_{i}\right| ^{2}=A_{i}A_{i}^{+}......(2)$ and $\left| y_{j}\right|
^{2}=B_{j}^{+}B_{j}......(3)$, $i,j=0,1,...,(n-1)$, respectively. Under the
supposition we can show that the above $x_{i}$ in (2) and $y_{j}$ in (3)
satisfy (1). Let us consider the case in which all the $a_{ij}$ are real. It
is not hard to extend the result to the case in which all the $a_{ij}$ are
complex. We only show $\left| x_{0}y_{0}\right| ^{2}=\left| a_{00}\right|
^{2}$ and omit the others. From (2) and (3), $\left| x_{0}y_{0}\right| ^{2}=$
$A_{0}A_{0}^{+}B_{0}^{+}B_{0}=(\sum_{j=0}^{n-1}\left| a_{0j}\right|
^{2})(\sum_{i=0}^{n-1}\left| a_{i0}\right| ^{2})=\sum_{i,j=0}^{n-1}\left|
a_{0j}\right| ^{2}\left| a_{i0}\right| ^{2}=\sum_{i,j=0}^{n-1}\left|
a_{00}\right| ^{2}\left| a_{ij}\right| ^{2}=\left| a_{00}\right| ^{2}$. In
the last but one step we use the equality $\left| a_{0j}\right| ^{2}\left|
a_{i0}\right| ^{2}=\left| a_{00}\right| ^{2}\left| a_{ij}\right| ^{2}$,
which holds since $\left(
\begin{tabular}{cc}
$a_{00}$ & $a_{0j}$ \\
$a_{i0}$ & $a_{ij}$%
\end{tabular}%
\right) $ is a $2\times 2$ submatrix of $M$. This completes the proof.

{\large Corollary}

If $|\psi \rangle $ is separable then $\det (M)=0$.

\section{The separability for a $n-$qubit system}

Let $|\psi \rangle $ be a pure state of a $n-$qubit system. Then we can
write $|\psi \rangle =\sum_{i_{1},i_{2},...,i_{n}\in
\{0,1\}}a_{i_{1}i_{2}...i_{n}}|i_{1}i_{2}...i_{n}\rangle $. Let the density
operator $\rho ^{12...n}=|\psi \rangle \langle \psi |$ and $\rho
^{12...(i-1)(i+1)...n}$ be the reduced density operator obtained by tracing
out the $ith$ qubit. Then $\rho ^{12...(i-1)(i+1)...n}=tr_{i}(\rho
^{12...n})=M_{i}M_{i}^{+}$, where $i=1,2,...,n$ and $M_{i}$ are $%
2^{(n-1)}\times 2$ matrices of the form $\left(
a_{b_{1}b_{2}...b_{i-1}0b_{i+1}...b_{n}},a_{b_{1}b_{2}...b_{i-1}1b_{i+1}...b_{n}}\right)
$ in which $b_{1}$,$b_{2}$,$...$,$b_{n}\in \{0,1\}$.

For example, let $|\psi \rangle $ be a state of a 3-qubit system. Then $%
|\psi \rangle $ can be written as $|\psi \rangle
=\sum_{i=0}^{7}a_{i}|i\rangle $. $M_{3}$ is a $4\times 2$ matrix $\left(
\begin{tabular}{cc}
$a_{0}$ & $a_{1}$ \\
$a_{2}$ & $a_{3}$ \\
$a_{4}$ & $a_{5}$ \\
$a_{6}$ & $a_{7}$%
\end{tabular}%
\right) $. Each entry of $M_{3}$ is an amplitude of $|\psi \rangle $.

There are three versions of the separability.

Version 1. $|\psi \rangle $ is separable if and only if the determinants of
all the $2\times 2$ submatrices of $M_{1}$,$M_{2}$,.... and $M_{n}$ are zero.

The proof of version 1 is similar to the one for a bipartite system in
section 2.

Version 2. $|\psi \rangle $ is separable if and only if $%
a_{i}a_{j}=a_{k}a_{l}$, where $i+j=k+l$ and $i\oplus j=k\oplus l$ where $%
0\leq i,j,k,l\leq 2^{n}-1$ are $n-$bit strings$\ $and $\oplus $ indicates
addition modulo 2.

For example, $2$, $7$, $5$ and $4$ can be written in binary numbers as $%
010,111,101$ and $100$, respectively. It is well known $010+111$(modulo 2)$%
=101$, $101$ $+$ $100=001$(modulo 2). Therefore $2+7\neq 5+4$(modulo 2)
though $2+7=5+4=9$.

Using this condition it is easy to verify that states $|W\rangle =1/\sqrt{n}%
(|2^{0}\rangle +$ $|2^{1}\rangle +...+|2^{n-1}\rangle $ and $|GHZ\rangle =1/%
\sqrt{2}(|0^{(n)}\rangle +|1^{(n)}\rangle )$ for a $n-$qubit system \cite%
{Dur}\ are entangled.

Let $i_{1}i_{2}...i_{n}$, $j_{1}j_{2}...j_{n}$, $k_{1}k_{2}...k_{n}$ and $%
l_{1}l_{2}...l_{n}$ be $n-$bit strings of $i$,$j,k$ and $l$, respectively.
Then version 3 is phrased below.

Version 3. $|\psi \rangle $ is separable if and only if $%
a_{i}a_{j}=a_{k}a_{l}$, where $\{i_{t},j_{t}\}=\{k_{t},l_{t}\}$, $%
t=1,2,...,n $.

The following lemma 1 shows that versions 2 and 3 are equivalent to each
other.

Lemma 1. $i+j=k+l$ and $i\oplus j=k\oplus l$ if and only if $%
\{i_{t},j_{t}\}=\{k_{t},l_{t}\}$, $t=1,2,...,n$.

The proof of lemma 1 is put in appendix A.

We argue version 3 next.

Assume that $|\psi \rangle =(x_{0}^{(1)}|0\rangle +x_{1}^{(1)}|1\rangle
)\otimes (x_{0}^{(2)}|0\rangle +x_{1}^{(2)}|1\rangle )\otimes ...\otimes
(x_{0}^{(n)}|0\rangle +x_{1}^{(n)}|1\rangle )$. By tensor product $%
x_{i_{1}}^{(1)}x_{i_{2}}^{(2)}....x_{i_{n}}^{(n)}=a_{i_{1}i_{2}...i_{n}}$,
where $i_{t}=0,1$, $t=1,2,...,n$. Then $%
a_{i}a_{j}=x_{i_{1}}^{(1)}x_{j_{1}}^{(1)}x_{i_{2}}^{(2)}x_{j_{2}}^{(2)}....x_{i_{n}}^{(n)}x_{j_{n}}^{(n)}
$ and $%
a_{k}a_{l}=x_{k_{1}}^{(1)}x_{l_{1}}^{(1)}x_{k_{2}}^{(2)}x_{l_{2}}^{(2)}....x_{k_{n}}^{(n)}x_{l_{n}}^{(n)}
$. Explicitly, $a_{i}a_{j}=a_{k}a_{l}$ whenever $\{i_{t}$,$%
j_{t}\}=\{k_{t},l_{t}\}$, $t=1,2,...,n$.

Conversely, suppose that $a_{i}a_{j}=a_{k}a_{l}$ whenever $%
\{i_{t},j_{t}\}=\{k_{t},l_{t}\}$, $t=1,2,...,n$. Let $\left|
x_{i_{t}}^{(t)}\right| ^{2}=\sum_{i_{1},..,i_{t-1},i_{t+1},..,i_{n}\in
\{0,1\}}\left| a_{i_{1}i_{2},...,i_{n}}\right| ^{2}$, where $t=1,2,...,n$.
We can show $%
|x_{i_{1}}^{(1)}x_{i_{2}}^{(2)}....x_{i_{n}}^{(n)}|^{2}=|a_{i_{1}i_{2}...i_{n}}|^{2}
$. We only demonstrate the cases of $n=2$ and $3$ to give the essential
ideas of the general case.

When $n=2$, see section 2. When $n=3$, see appendix $B$. The two cases
suggest that it be simpler to prove $%
|x_{i_{1}}^{(1)}x_{i_{2}}^{(2)}....x_{i_{n}}^{(n)}|^{2}=|a_{i_{1}i_{2}...i_{n}}|^{2}\left( \sum |a_{i_{1}i_{2}...i_{n}}|^{2}\right) ^{n-1}
$. Now we finish the argument for the real number case. It is not hard to
extend the result to the complex number case.

\section{The separability for a multi ($n$)$-$partite system}

Assume that each subsystem has the same dimension $d$. Let $|i_{t}\rangle $
be the orthonormal basis $|0\rangle $,$|1\rangle $,...,$|(d-1)\rangle $ for
the $tth$ subsystem. Then any pure state $|\psi \rangle $\ can be written as
$|\psi \rangle
=\sum_{i_{1},i_{2},...,i_{n}=0}^{d-1}a_{i_{1}i_{2}...i_{n}}|i_{1}i_{2}...i_{n}\rangle
$. Assume that $|\psi \rangle $ is separable. Then we can write $|\psi
\rangle =\left( \sum_{i_{1}=0}^{d-1}x_{i_{1}}^{(1)}|i_{1}\rangle \right)
\otimes \left( \sum_{i_{2}=0}^{d-1}x_{i_{2}}^{(2)}|i_{2}\rangle \right)
\otimes ...\otimes \left( \sum_{i_{n}=0}^{d-1}x_{i_{n}}^{(n)}|i_{n}\rangle
\right) $. By tensor product $%
x_{i_{1}}^{(1)}x_{i_{2}}^{(2)}....x_{i_{n}}^{(n)}=a_{i_{1}i_{2}...i_{n}}$,
where $i_{1}$,$i_{2}$,$...$,$i_{n}\in \{0,1,...,(d-1)\}$.

Let the density operator $\rho ^{12...n}=|\psi \rangle \langle \psi |$ and $%
\rho ^{12...(i-1)(i+1)...n}$ be the reduced density operator obtained by
tracing out the $ith$ subsystem. Then $\rho
^{12...(i-1)(i+1)...n}=tr_{i}(\rho ^{12...n})=M_{i}M_{i}^{+}$, where $%
i=1,2,...,n$ and $M_{i}$ are $d^{n-1}\times d$ matrices of the amplitudes of
the form

\noindent $\left(
a_{k_{1}k_{2}...k_{i-1}0k_{i+1}...k_{n}},a_{k_{1}k_{2}...k_{i-1}1k_{i+1}...k_{n}},...,a_{k_{1}k_{2}...k_{i-1}(d-1)k_{i+1}...k_{n}}\right)
$, where $k_{1}$,$k_{2}$,...,$k_{i-1}$,$k_{i+1}$,...,$k_{n}\in
\{0,1,...,(d-1)\}$.

There are two versions of the separability.

Version 1. $|\psi \rangle $ is separable if and only if the determinants of
all the $2\times 2$ submatrices of $M_{1}$, $M_{2}$, ... and $M_{n}$ are
zero.

Version 2. $|\psi \rangle $ is separable if and only if $%
a_{i_{1}i_{2}...i_{n}}a_{j_{1}j_{2}...j_{n}}=a_{k_{1}k_{2}...k_{n}}a_{l_{1}l_{2}...l_{n}}
$, where $\{i_{t},j_{t}\}=\{k_{t},l_{t}\}$, $t=1,2,...,n$.

The proof of version 1 is similar to the one for a bipartite system. The
proof of version 2 is similar to the one for a $n-$qubit system.

When $n=2$, the criterion is reduced to the one for a bipartite system. When
$d=2$, the criterion is reduced to the one for a $n$-qubit system.

\section{A necessary condition for a local unitary equivalence class for a
bipartite system}

We use the following lemma 2 to establish the necessary condition.

Lemma 2. Let $|\psi \rangle $ be a pure state of a composite system $AB$
possessed by both Alice and Bob. Let $M$ $=(a_{jk})_{n\times n}$ be the
matrix of the amplitudes of $|\psi \rangle $. Let $\rho ^{AB}=$ $|\psi
\rangle \langle \psi |$ and $\rho ^{A}$ $=tr_{B}(\rho ^{AB})$. Then $\left|
\det (M)\right| ^{2}$ is just the product of the eigenvalues of $\rho ^{A}$.

The proof is put in appendix C.

Lemma 2 reveals the relation between the determinant of the matrix of the
amplitudes and the eigenvalues of $\rho ^{A}$ for a bipartite system.

{\large The corollary of lemma 2}

Let $M_{\psi }$ ($M_{\phi }$) be the matrix of the amplitudes of a pure
state $|\psi \rangle $ ($|\phi \rangle $) of a composite system $AB$. Then $%
\left| \det (M_{\psi })\right| =\left| \det (M_{\phi })\right| $ whenever $%
|\psi \rangle $ $\sim $ $|\phi \rangle $. That is, $\left| \det (M_{\psi
})\right| $ is invariant under local unitary operators.

It is well known that it only needs $O(n^{3})$ multiplication operations to
compute $\left| \det (M)\right| $ instead of doing Schmidt decomposition in %
\cite{Nielsen99}\cite{Peres}.

For a two-qubit system, let $|\psi \rangle =a|00\rangle +b|01\rangle
+c|10\rangle +d|11\rangle $ and $\rho ^{12}=|\psi \rangle \langle \psi |$.
By lemma 2 $|ad-bc|^{2}$ is the product of the eigenvalues of $\rho ^{1}$.
Let $|ad-bc|=\in $. We can show that $\in $ satisfies $0\leq \in \leq \frac{1%
}{2}$ and eigenvalues $\lambda _{\pm }=\frac{1\pm \sqrt{1-4\in ^{2}}}{2}$.
Hence, $|\psi \rangle \sim \sqrt{\lambda _{+}}|00\rangle +\sqrt{\lambda _{-}}%
|11\rangle $ or $|\psi \rangle \sim \sqrt{\lambda _{-}}|00\rangle +\sqrt{%
\lambda _{+}}|11\rangle $.\

\section{The variance as a measure of entanglement}

We obtain the necessary and sufficient conditions of separability in
sections 2, 3 and 4. Apparently, $\left|
a_{i_{1}i_{2}...i_{n}}a_{j_{1}j_{2}...j_{n}}-a_{k_{1}k_{2}...k_{n}}a_{l_{1}l_{2}...l_{n}}\right|
$, where $\{i_{t},j_{t}\}=\{k_{t},l_{t}\}$, $t=1,2,...,n$., is just a
deviation from a product state. It is intuitive to suggest the variance: $%
\sum \left|
a_{i_{1}i_{2}...i_{n}}a_{j_{1}j_{2}...j_{n}}-a_{k_{1}k_{2}...k_{n}}a_{l_{1}l_{2}...l_{n}}\right| ^{2}
$, where $\{i_{t},j_{t}\}=\{k_{t},l_{t}\}$, $t=1,2,...,n$, as a measure of
entanglement of $|\psi \rangle $. Let $D_{E}(|\psi \rangle )$ be the measure
of entanglement.

$D_{E}(|\psi \rangle )$ has the following properties.

Property 1. $D_{E}(|\psi \rangle )=0$ if and only if $|\psi \rangle $ is
separable.

{\large The properties for a two-qubit system}

For a two-qubit system, let $|\psi \rangle =a|00\rangle +b|01\rangle
+c|10\rangle +d|11\rangle $. Then $D_{E}(|\psi \rangle )=\left| ad-bc\right|
^{2}$.

Property 2. The maximum of $D_{E}(|\psi \rangle )=\left| ad-bc\right|
^{2}\leq (\left| ad\right| +\left| bc\right| )^{2}\leq (\frac{\left|
a\right| ^{2}+\left| d\right| ^{2}}{2}+\frac{\left| b\right| ^{2}+\left|
c\right| ^{2}}{2})^{2}=\frac{1}{4}.$

When $a,b,c$ and $d$ are real, by computing extremum it is derived that the
maximally entangled states must be of the forms: $x|00\rangle +y|01\rangle
-y|10\rangle +x|11\rangle $ or $x|00\rangle +y|01\rangle +y|10\rangle
-x|11\rangle $.

Property 3. $|\psi \rangle $ $\sim $ $|\psi ^{\prime }\rangle $ if and only
if $D_{E}(|\psi \rangle )=D_{E}(|\psi ^{\prime }\rangle )$.

Given $|\psi \rangle =a|00\rangle +b|01\rangle +c|10\rangle +d|11\rangle $
and $|\psi ^{\prime }\rangle =a^{\prime }|00\rangle +b^{\prime }|01\rangle
+c^{\prime }|10\rangle +d^{\prime }|11\rangle $. Suppose that $|\psi \rangle
\sim |\psi ^{\prime }\rangle $. By the necessary condition in section 5, $%
D_{E}(|\psi \rangle )=D_{E}(|\psi ^{\prime }\rangle )$.

Conversely, suppose $D_{E}(|\psi \rangle )=D_{E}(|\psi ^{\prime }\rangle )$.
Let us show $|\psi \rangle \sim |\psi ^{\prime }\rangle $. Using Schmidt
decomposition, we can write $|\psi \rangle \sim \sqrt{\lambda _{1}}%
|00\rangle +\sqrt{\lambda _{2}}|11\rangle $, where $\lambda _{1}+\lambda
_{2}=1$. As discussed above $|ad-bc|=\sqrt{\lambda _{1}}\sqrt{\lambda _{2}}$%
. As well using Schmidt decomposition\ we can write $|\psi ^{\prime }\rangle
\sim \sqrt{\rho _{1}}|00\rangle +\sqrt{\rho _{2}}|11\rangle $, where $\rho
_{1}+\rho _{2}=1$, and $|a^{\prime }d^{\prime }-b^{\prime }c^{\prime }|=%
\sqrt{\rho _{1}}\sqrt{\rho _{2}}$. Thus $\lambda _{1}\lambda _{2}=\rho
_{1}\rho _{2}$. Then $\lambda _{1}(1-\lambda _{1})=\rho _{1}(1-\rho _{1})$.
There are two cases. Case 1. $\lambda _{1}=\rho _{1}$. Then $\lambda
_{2}=\rho _{2}$. Case 2. $\lambda _{1}+\rho _{1}+1=0$. In the case $\lambda
_{2}=\rho _{1}$ and $\lambda _{1}=\rho _{2}$. It means that $|\psi \rangle $
and $|\psi ^{\prime }\rangle $ have the same Schmidt co-efficient for either
of the two cases. By factor 5 in \cite{Nielsen99}\cite{Peres}, $|\psi
\rangle \sim |\psi ^{\prime }\rangle $.

Nielsen in \cite{Nielsen99} showed $|\psi ^{\prime }\rangle $ $\sim $ $|\psi
^{\prime \prime }\rangle $ by calculating eigenvalue, where $|\psi ^{\prime
}\rangle =\sqrt{\alpha _{+}}|00\rangle +\sqrt{\alpha _{-}}|11\rangle $, and $%
|\psi ^{\prime \prime }\rangle =(|00\rangle +|1\rangle (\cos \gamma
|0\rangle +\sin \gamma |1\rangle ))/\sqrt{2}$. By property 3 it only needs
to check $\sqrt{\alpha _{+}}\sqrt{\alpha _{-}}=\sin \gamma /2$.

{\large Conclusion}

In this paper we have presented the necessary and sufficient
conditions of separability for multipartite pure states. These
conditions don't require Schmidt decomposition or tracing out
operations. By using the conditions it is easy to check whether or
not a multipartite pure state is entangled.


Appendix A. The proof of lemma 1

Let $\alpha _{1}\alpha _{2}...\alpha _{n}$, $\beta _{1}\beta _{2}...\beta
_{n}$, $\delta _{1}\delta _{2}...\delta _{n}$ and $\gamma _{1}\gamma
_{2}...\gamma _{n}$ be the $n-$bit strings of $\alpha $, $\beta $, $\delta $
and $\gamma $, respectively.

Lemma 1. \{$\alpha _{i}$, $\beta _{i}$\}=\{$\delta _{i}$, $\gamma _{i}$\}$,$
$i=1,2,...,n$, if and only if $\alpha +\beta =\delta $ $+$ $\gamma $ and $%
\alpha \oplus \beta =\delta $ $\oplus $ $\gamma ,$ where $\oplus $ indicates
addition modulo 2.

Proof. Suppose \{$\alpha _{i}$, $\beta _{i}$\}=\{$\delta _{i}$, $\gamma _{i}$%
\}$,$ $i=1,2,...,n$. Since $\alpha +\beta =(\alpha _{1}+\beta
_{1})2^{n-1}+(\alpha _{2}+\beta _{2})2^{n-2}+...+(\alpha _{n}+\beta _{n})$
and $\delta $ $+$ $\gamma =(\delta _{1}+\gamma _{1})2^{n-1}+(\delta
_{2}+\gamma _{2})2^{n-2}+...+(\delta _{n}+\gamma _{n})$, by the supposition
it is easy to see $\alpha +\beta =\delta $ $+$ $\gamma $. It is
straightforward to obtain $\alpha _{1}\alpha _{2}...\alpha _{n}\oplus \beta
_{1}\beta _{2}...\beta _{n}=\delta _{1}\delta _{2}...\delta _{n}\oplus
\gamma _{1}\gamma _{2}...\gamma _{n}$.

Conversely, suppose $\alpha +\beta =\delta $ $+$ $\gamma $ and $\alpha
\oplus \beta =\delta $ $\oplus $ $\gamma $. First let us consider the case
where $n=1$. There are three cases.

Case 1. $\alpha _{1}+\beta _{1}=\delta _{1}+\gamma _{1}=0$. This means $%
\alpha _{1}=\beta _{1}=\delta _{1}=\gamma _{1}=0$.

Case 2. $\alpha _{1}+\beta _{1}=\delta _{1}+\gamma _{1}=1$. This implies $%
\{\alpha _{1},\beta _{1}\}=\{\delta _{1},\gamma _{1}\}=\{1,0\}$.

Case 3. $\alpha _{1}+\beta _{1}=\delta _{1}+\gamma _{1}=2$. \ This says $%
\alpha _{1}=\beta _{1}=\delta _{1}=\gamma _{1}=1$.

No matter which of the above three cases happens, it yields \{$\alpha _{1}$,
$\beta _{1}$\}=\{$\delta _{1}$, $\gamma _{1}$\}.

Let us consider the case $n$. Since $\alpha +\beta =\delta $ $+$ $\gamma $, $%
(\alpha _{1}+\beta _{1})2^{n-1}+(\alpha _{2}+\beta _{2})2^{n-2}+...+(\alpha
_{n}+\beta _{n})=(\delta _{1}+\gamma _{1})2^{n-1}+(\delta _{2}+\gamma
_{2})2^{n-2}+...+(\delta _{n}+\gamma _{n})$. Again since $\alpha \oplus
\beta =\delta $ $\oplus $ $\gamma $, that is, $\alpha _{1}\alpha
_{2}...\alpha _{n}\oplus \beta _{1}\beta _{2}...\beta _{n}=\delta _{1}\delta
_{2}...\delta _{n}\oplus \gamma _{1}\gamma _{2}...\gamma _{n}$, we obtain $%
\alpha _{i}\oplus \beta _{i}=\delta _{i}\oplus \gamma _{i}$, $i=1,2,...,n$.
\ There are two cases.

Case 1. $\alpha _{n}\oplus \beta _{n}=\delta _{n}\oplus \gamma _{n}=1$. In
the case $\{\alpha _{n}$, $\beta _{n}\}=\{\delta _{n}$, $\gamma
_{n}\}=\{0,1\}$. Then $(\alpha _{1}+\beta _{1})2^{n-2}+(\alpha _{2}+\beta
_{2})2^{n-3}+...+(\alpha _{n-1}+\beta _{n-1})=(\delta _{1}+\gamma
_{1})2^{n-2}+(\delta _{2}+\gamma _{2})2^{n-3}+...+(\delta _{n-1}+\gamma
_{n-1})$ \ \ and $\alpha _{i}\oplus \beta _{i}=\delta _{i}\oplus \gamma _{i}
$, $i=1,2,...,n-1$. By induction hypothesis $\{\alpha _{i},\beta
_{i}\}=\{\delta _{i},\gamma _{i}\}$, $i=1,2,...,n-1$.

Case 2. $\alpha _{n}\oplus \beta _{n}=\delta _{n}\oplus \gamma _{n}=0$.
There are two subcases.

Subcase 2.1. $\alpha _{n}=\beta _{n}=\delta _{n}=\gamma _{n}=0$ or $\alpha
_{n}=\beta _{n}=\delta _{n}=\gamma _{n}=1$. As discussed in case 1, we can
obtain $\{\alpha _{i},\beta _{i}\}=\{\delta _{i},\gamma _{i}\}$, $%
i=1,2,...,n-1$ by induction hypothesis.

Subcase 2.2. $\alpha _{n}=\beta _{n}=1$ and $\delta _{n}=\gamma _{n}=0$ or $%
\alpha _{n}=\beta _{n}=0$ and $\delta _{n}=\gamma _{n}=1$. Let us consider
the former case. In the case $(\alpha _{1}+\beta _{1})2^{n-2}+(\alpha
_{2}+\beta _{2})2^{n-3}+...+(\alpha _{n-2}+\beta _{n-2})2+(\alpha
_{n-1}+\beta _{n-1}+1)=$

$(\delta _{1}+\gamma _{1})2^{n-2}+(\delta _{2}+\gamma
_{2})2^{n-3}+...+(\delta _{n-2}+\gamma _{n-2})2+(\delta _{n-1}+\gamma
_{n-1}) $.

Since $\alpha _{n-1}\oplus \beta _{n-1}=\delta _{n-1}\oplus \gamma _{n-1}$,
either $\alpha _{n-1}\oplus \beta _{n-1}=\delta _{n-1}\oplus \gamma _{n-1}=0$
or $1$ causes that one of $(\alpha _{n-1}+\beta _{n-1}+1)$ and $(\delta
_{n-1}+\gamma _{n-1})$ is odd and the other is even. It contradicts $\alpha
\oplus \beta =\delta $ $\oplus $ $\gamma $. \

Appendix B. The separability for a $n-$qubit system

When $n=3,$\ let us show $%
|x_{i_{1}}^{(1)}x_{i_{2}}^{(2)}x_{i_{3}}^{(3)}|^{2}=|a_{i_{1}i_{2}i_{3}}|^{2}
$ when $a_{i}a_{j}=a_{k}a_{l}$, where $\{i_{t},j_{t}\}=\{k_{t},l_{t}\}$, $%
t=1,2,3$. We only illustrate $%
|x_{0}^{(1)}x_{0}^{(2)}x_{0}^{(3)}|^{2}=|a_{000}|^{2}$. Other cases then
follow readily. Experientially, it is simpler to prove $%
|x_{0}^{(1)}x_{0}^{(2)}x_{0}^{(3)}|^{2}$ $=|a_{000}|^{2}(\sum_{i,j,k\in
\{0,1\}}|a_{ijk}|^{2})(\sum_{i,j,k\in \{0,1\}}|a_{ijk}|^{2})$, where $%
|x_{0}^{(1)}|^{2}=\sum_{i,j\in \{0,1\}}|a_{0ij}|^{2}$, $|x_{0}^{(2)}|^{2}=%
\sum_{k,l\in \{0,1\}}|a_{k0l}|^{2}$ and $|x_{0}^{(3)}|^{2}=\sum_{p,q\in
\{0,1\}}|a_{pq0}|^{2}$.

First we show that $a_{0ij}a_{k0l}a_{pq0}$ can be rewritten as $%
a_{000}a_{\alpha _{1}\alpha _{2}\alpha _{3}}a_{\delta _{1}\delta _{2}\delta
_{3}}$. There are the following four cases.

Case 1. Consider $a_{0ij}a_{k0l}$ and the pairs $\{0,k\},\{i,0\}$ and $%
\{j,l\}$. If $j\ast l=0$ , then $a_{0ij}a_{k0l}=a_{000}a_{ki(j+l)}$ since $%
\{j,l\}=\{0,j+l\}$.

Case 2. Consider $a_{0ij}a_{pq0}$ and the pairs $\{0,p\},\{i,q\}$ and $%
\{j,0\}$. If $i\ast q=0$, then $a_{0ij}a_{pq0}=a_{000}a_{p(i+q)j}$ since $%
\{i,q\}=\{0,i+q\}$.

Case 3. Consider $a_{k0l}a_{pq0}$ and the pairs $\{k,p\},\{0,q\}$ and $%
\{l,0\}$. If $k\ast p=0$, then $a_{k0l}a_{pq0}=a_{000}a_{(k+p)ql}$ since $%
\{k,p\}=\{0,k+p\}$.

Case 4. Otherwise $i=j=l=k=p=q=1$. It is not hard to derive $%
a_{3}a_{5}a_{6}=a_{1}a_{7}a_{6}=a_{0}a_{7}^{2}$.

Second, let us show that $a_{000}a_{\alpha _{1}\alpha _{2}\alpha
_{3}}a_{\delta _{1}\delta _{2}\delta _{3}}$\ can be rewritten\ as $%
a_{0ij}a_{k0l}a_{pq0}$. If $a_{000}a_{\alpha _{1}\alpha _{2}\alpha
_{3}}a_{\delta _{1}\delta _{2}\delta _{3}}$ is of the forms: $%
a_{000}a_{0ij}a_{k0l}$,\ $a_{000}a_{0ij}a_{pq0}$ or $a_{000}a_{k0l}a_{pq0}$,
then these forms are desired. Otherwise $a_{000}a_{\alpha _{1}\alpha
_{2}\alpha _{3}}a_{\delta _{1}\delta _{2}\delta _{3}}$\ must be $%
a_{0}a_{6}a_{6}$, $a_{0}a_{3}a_{3},$ $a_{0}a_{5}a_{5}$ or\ of the form $%
a_{0}a_{7}a_{rst}$, which can be rewritten as $a_{2}a_{4}a_{6}$, $%
a_{1}a_{2}a_{3}$, $a_{1}a_{4}a_{5}$, $a_{1}a_{6}a_{rst}$, respectively. $%
a_{2}a_{4}a_{6}$, $a_{1}a_{2}a_{3}$ and $a_{1}a_{4}a_{5}$ are just desired
and $a_{1}a_{6}a_{rst}$ is furthermore rewritten as follows. There are three
cases.

Case 1. In the case $r=0$ or $s=0,$ this is desired.

Case 2. In the case $r=s=t=1$, $a_{1}a_{6}a_{7}=a_{3}a_{5}a_{6},$ desired.

Case 3. In the case $r=s=1$ and $t=0$, $a_{1}a_{6}a_{6}=a_{2}a_{5}a_{6}$,
desired.

Appendix C. The proof of lemma 2

Proof. Suppose that systems $A$ and $B$ have the same dimensions $n$. Let $%
|\psi \rangle =\sum_{i,j=0}^{n-1}a_{ij}|i\rangle |j\rangle $. Then $%
M=(a_{ij})_{n\times n}$. Let density operator $\rho ^{AB}=|\psi \rangle
\langle \psi |$. Then $\rho ^{AB}=(\sum_{i,j=0}^{n-1}a_{ij}|i\rangle
|j\rangle )(\sum_{l,k=0}^{n-1}a_{lk}^{\ast }\langle l|\langle
k|)=\sum_{i,j=0}^{n-1}\sum_{l,k=0}^{n-1}a_{ij}a_{lk}^{\ast }|i\rangle
|j\rangle \langle l|\langle k|$

$=\sum_{i,l=0}^{n-1}\sum_{j,k=0}^{n-1}a_{ij}a_{lk}^{\ast }|i\rangle
|j\rangle \langle l|\langle k|$. The reduced density operator for system $A$
is defined by $\rho ^{A}=tr_{B}(\rho ^{AB})$. Let us compute $\rho ^{A}$.

$\rho ^{A}=\sum_{i,l=0}^{n-1}\sum_{j,k=0}^{n-1}a_{ij}a_{lk}^{\ast }|i\rangle
\langle l|\delta _{kj}$ (where $\delta _{kj}=1$ when $k=j$. Otherwise $0$.) $%
=\sum_{i,l=0}^{n-1}\sum_{j=0}^{n-1}a_{ij}a_{lj}^{\ast }|i\rangle \langle
l|=\sum_{i,l=0}^{n-1}(\sum_{j=0}^{n-1}a_{ij}a_{lj}^{\ast })|i\rangle \langle
l|$. Let $A_{i}=(a_{i0},a_{i1},....a_{i(n-1)})$, that is, the $ith$ row of $%
A $. Then $\sum_{j=0}^{n-1}a_{ij}a_{lj}^{\ast }=A_{i}A_{l}^{+}$. Finally $%
\rho ^{A}=\sum_{i,l=0}^{n-1}A_{i}A_{l}^{+}|i\rangle \langle l|=\left(
\begin{tabular}{c}
$A_{0}$ \\
$A_{1}$ \\
$\vdots $ \\
$A_{n-1}$%
\end{tabular}%
\right) (A_{0}^{+},A_{1}^{+},...,A_{n-1}^{+})=MM^{+}$. Thus $\det (\rho
^{A})=|\det (M)|^{2}$. Hence $\left| \det (M)\right| ^{2}$ is just the
product of the eigenvalues of $\rho ^{A}$. Q.E.D.


\begin{thebibliography}{99}
\bibitem{Nielsen99} M.A. Nielsen, Phys. Rev. Lett. 83, 436(1999).

\bibitem{Peres} A. Peres, Quantum theory: Concepts and methods (Kluwer
Academic Dordrecht, 1993). P. 123.

\bibitem{Nielsen00} M. A. Nielsen and I. L. Chuang, Quantum computation and
quantum information (Cambridge University Press, Cambridge, England, 2000).
p. 109.

\bibitem{Peres95} A. Peres, Phys. Lett. A 202, 16 (1995).

\bibitem{Peres96} A. Peres, Phys. Rev. Lett. 77, 1413 (1996).

\bibitem{Thapliyal99} A. V. Thapliyal, Phys. Rev. A 59, 3336 (1999).

\bibitem{Grassl} M. Grassl et al., Phys. Rew. A. 58 (1998) 1833-1839.

\bibitem{Sudbery00} H. A. Carteret and A. Sudbery, J. Phys. A: Math. Gen. 33
(2000)4981-5002.

\bibitem{Sudbery01} A. Sudbery, J. Phys. A: Math. Gen. 34 (2001)643-652.

\bibitem{Acin} A. Acin et al., Phys. Rew. Lett. 85 (2000) 1560-1563.

\bibitem{Bennett00} C. H. Bennett et al., Phys. Rev. A., 63(2000) 012307.

\bibitem{Meyer} D. A. Meyer and N. R. Wallach, J. of mathematical physics 43
(2002) 4273-4278.

\bibitem{Dur} W. D$\ddot{u}$r et al., Phys. Rev. A., 62(2000)062314.
\end{thebibliography}
\end{document}